\documentclass[twoside]{ilcws10}
\usepackage[latin1]{inputenc}
\usepackage{graphicx,color}
\usepackage{amssymb,amsmath,array}

\begin{document}
\title{
Next Generation CALICE Electromagnetic Calorimeter
} 
\author{Denis Grondin$^1$ and Daniel Jeans$^2$ {\em on behalf of the CALICE Si-W ECAL group}
\vspace{.3cm}\\
1- LPSC - UJF - IN2P3/CNRS - INPG - Grenoble - France
\vspace{.1cm}\\
2- Laboratoire Leprince-Ringuet -  \'Ecole polytechnique, CNRS/IN2P3 - Palaiseau - France
}


\maketitle

\begin{abstract}

This paper presents mechanical R\&D for the CALICE Silicon-tungsten electromagnetic calorimeter. 
After the physics ECAL prototype, tested in 2006 (DESY-CERN), 2007 (CERN), 2008 (FNAL) and before 
the design of different ``modules 0'' (barrel and end-cap) for a final detector, a technological ECAL prototype, 
called the EUDET module, is under design in order to have a close to full scale technological solution which could 
be used for the final detector, taking into account future industrialisation of production.

\end{abstract}

\section{Introduction}

The electromagnetic calorimeter has been optimized for the reconstruction of photons and electrons 
and for separating them properly from debris coming from charged hadron interactions in the device. 
The detector is designed to be used in particle flow--based event reconstruction. It is conceived
as a sampling calorimeter with tungsten absorber (to reduce the Moliere radius of EM showers, 
longitudinally separate hadronic and EM showers, and reduce the total detector thickness) together
with silicon-based active layers, chosen to limit the detector thickness, allow highly granular
readout, and stable operation with respect to environmental conditions.
The following sampling is then under consideration: 20 layers of 0.6 X0 thick tungsten 
absorbers (2.1 mm) and another 9 layers of tungsten 1.2 X0 thick (4.2 mm). 

\section{ILD Electromagnetic Si-W Calorimeter design}

The ILD design, with a solenoid outside the calorimeter, imposes an overall cylindrical symmetry. 
The global design has been developed with an attempt to simplify the device as much as possible, 
by reducing the number of different module types and different technologies used.

\subsection{Barrel geometry}

\begin{figure}
\centerline{
\includegraphics[width=0.45\columnwidth]{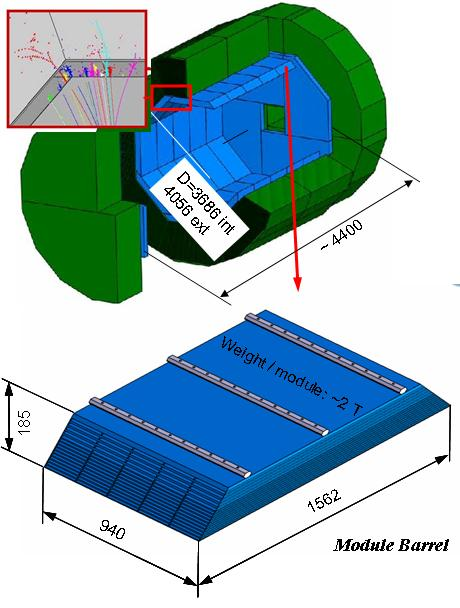}
\includegraphics[width=0.45\columnwidth]{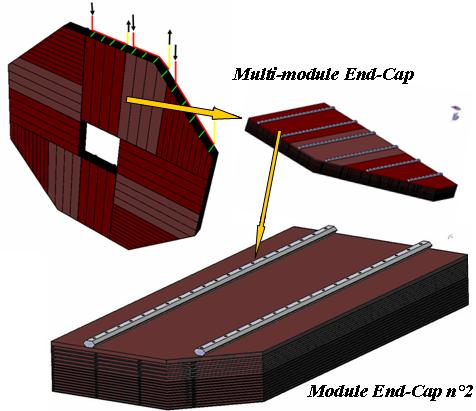}
}
\caption{General design of the calorimeter system, and 
the structure of modules of the ECAL barrel and endcap}
\label{fig:barrel_endcap}
\end{figure}

One of the requirements for the calorimeter is to ensure the best possible hermeticity. To minimize the 
number and effect of cracks in the barrel, a design with large modules is preferred, with boundaries not 
pointing to the vertex. As shown in Figure \ref{fig:barrel_endcap} the perfect $\phi$ symmetry of the coil has been approximated by 
an eight-fold symmetry and the 40 trapezoidal modules are installed in such a way that the cracks are at 
a very large angle with respect to the radial direction (trapezoidal shape). This octagonal shape seems 
to optimize the barrel modules' size and their mechanical properties without diverging too far from a circle. 
One eighth of the barrel calorimeter is called a stave. At the back of a stave, between the ECAL and the HCAL, 
some space is left which is used to house different services like cooling or electrical power and signal 
distribution. Along the beam axis, a stave is subdivided into five modules.

\subsection{End-cap geometry}

The ECAL end caps could be constructed from very similar modules but with different shape. 
To ensure that the depth of the calorimeter remains sufficient, the octagonal shape of the end-cap at the 
outer radius (2093 mm) follows the barrel part (1770 mm). Each end-cap consists of twelve modules 
(4x3 distinct types). The longest alveoli will be 2.50 m long. With this design, no crack is pointing to the origin. 

\subsection{Active layers}

These alveolar structures are used to house the active detection layers. A ``h'' structure consists of a central layer of tungsten
wrapped in carbon fibre composite, sandwiched between two detection slabs. These detection slabs are made up of a string of
Active Sensor Units (ASU) which consist of a PCB with embedded FE ASICs, mounted onto the silicon sensors.
A $\sim 500 \mu m$ copper plate attached to the PCB layer equalises the temperature along the length of the slab.

The active layers of the calorimeter are based on matrices of $5\times5 {\rm mm^2}$ PIN diodes in $300 \mu m$
thick high resistivity silicon. This gives a good S/N ratio at the MIP level, and provides the high granularity
required for PFA performance.

Low power FE electronics, running in power pulsed mode to make use of the ILC beam structure, should dissipate
around $25 \mu W$ per channel. The thin copper sheet is sufficient to drain this to the detector edge, where it is 
evacuated by means of a leak-less cooling system.

\section{The EUDET prototype}

\begin{figure}
\centerline{
\includegraphics[width=0.45\columnwidth]{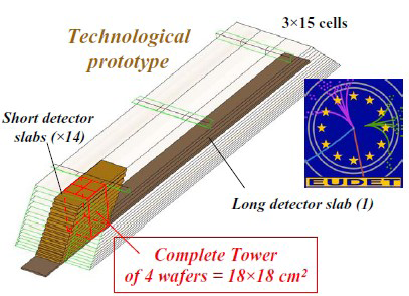}
\includegraphics[width=0.45\columnwidth]{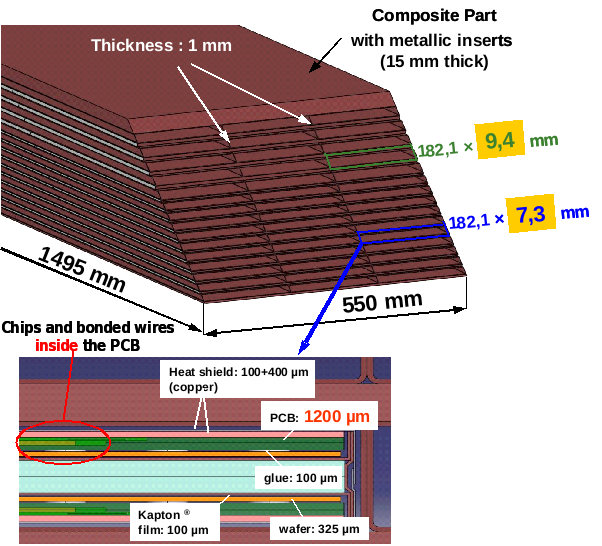}
}
\caption{Design of EUDET module}
\label{fig:eudet}
\end{figure}

The EUDET prototype (shown in Figure \ref{fig:eudet}) will be a proof of principle for the technological realisation 
of a SiW ECAL. It consists of a mechanical structure close to a barrel module of the 
final detector, and will be partially instrumented with silicon sensors, readout out
by front end electronics integrated into the detector volume, as is foreseen for 
the final detector design.
It is foreseen to equip a $18\times18 {\rm cm^2}$ tower of the EUDET mechanical structure, as well as one long
detection slab to test the readout of sensors at long distances.

\subsection{Alveolar structure}

The design and construction of a module presents an interesting technological challenge.
A design has been adopted where half of the tungsten layers are part of a solid mechanical structure, 
by embedding them into a light composite structure made of carbon fiber reinforced with epoxy.  
In between these plates and the carbon fiber partitions, free spaces are left into 
which detector elements, called "detector slabs", are inserted. This design has been validated on 
the physics prototype~\cite{caliceEcalComm}.

The method of an ``assembled structure'' has been chosen. Each alveolar layer is made independently, 
cut to the right length (at an angle of $45^o$) and assembled with the tungsten plates in a second curing step. 
This procedure limits the risks of wasting tungsten plates compared to an ``one block'' solution where all 
the structure is made in one step. After curing, the place holders (cores) for the detector slabs are removed, 
leaving empty spaces (called alveoli). This principle also reduces the cost of the industrial process: 
simpler moulds (one for barrel + 10 for end-caps) and the final  
piece is obtained in 2 simple polymerization processes, avoiding curing problems like thermal inertia, 
weight of metal mould, or the control of curing parameters. 

\subsubsection{Demonstrator prototype}

The first samples have been used to study mechanical behavior (destroying tests, dimensional controls...). 
The mechanical strength of glued structures has been validated for these multi-curing steps to obtain the 
final structure (weight 650kg for the EUDET module, up to 2T for the End-cap modules).
A first demonstrator of alveolar structure (1500 mm long) has been built mid 2008 to understand all 
manufacturing processes. Consisted of 3 alveolar layers + 2 Tungsten layers, 3 columns of cells, 
with a width based on physics prototype and an identical global length (1.3m and trapezoidal shape), 
this demonstrator has been used for thermal PCB studies and for cooling system analysis. 
This module has been equipped with a Bragg grating equipped optical fibre, allowing tests of
its structural characteristics.

\subsection{Sensors}

The development of sensors for the ECAL is focused on the improvement of their design, and on the reduction
of their cost. $9\times9 {\rm cm^2}$ sensors cut from larger wafers (than used for the physics prototype) are
now used, and are segmented into $5\times5 {\rm mm^2}$ readout cells. The edge of the sensor contains a guard ring,
preventing the breakdown of the sensor. Improvements to its design (reducing its width, and the coupling to
the detection pixels) are under study. A variety of industrial contacts are in the process of being established,
which will help to better understand the final cost, as well as strategies to reduce the cost.

\begin{figure}
\centerline{
\includegraphics[width=0.45\columnwidth]{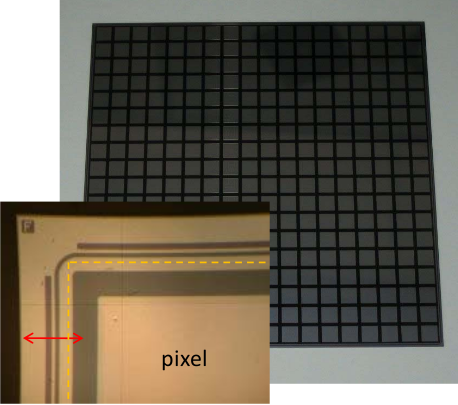}
\includegraphics[width=0.45\columnwidth]{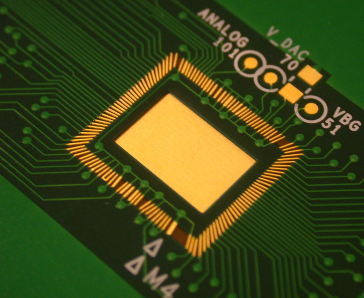}
}
\caption{Silicon detector and close-up of guard ring (left); well for bonding of ASIC in PCB (right)}
\label{fig:si_pcb}
\end{figure}

\subsection{FE electronics}

The FE electronics will be housed within the detector volume. The SKIROC2 chip is at present (June 2010) in fabrication,
and will soon be delivered. This 64-channel chip is designed to readout an ILD silicon-based ECAL, 
with a high dynamic range, and very low power consumption thanks to its power-pulsing capability. More details
are available in~\cite{caliceElecs}.
For tests in the meantime, the existing SPIROC2 chip
can be run in a SKIROC-emulating mode, albeit with a smaller number of channels and a reduced dynamic range,
which will allow first system tests to be undertaken.

\subsection{Detector integration}

The silicon sensors and FE ASICs will be mounted on a printed circuit board (PCB), whose thickness must be minimised in order
to reduce the insensitive volume of the detector. A thickness of under 1mm seems feasible.
This PCB is under development. Several prototypes have been produced, 
to test mechanical properties, signal routing, and the bonding of ASICs.

The unit composed of a single ($18\times18 {\rm cm^2}$) PCB, the silicon sensors and the ASICs is known as an 
Active Sensor Unit (ASU). Several of these ASUs must be joined together to make a full length detector slab.
This approach means that each detector unit can be tested before integration into the detector slabs.
Techniques for the manipulation and bonding of ASUs into complete detector slabs (incorporating HV supply,
heat extraction, mechanical strength) are under study, and show promising results. A 1m long slab of test boards 
has successfully been assembled.

\subsection{Cooling system and power dissipation}

Power dissipation is the key issue for this electronics. Indeed, the front-end electronics is located 
inside the calorimeter and has to have a very low consumption. The first slab thermal analysis is encouraging: 
assuming that the chip power is $25 \mu W$ per channel (power dissipation for the overall electronic chain including 
the digitization), a simulation of heat conduction just by the heat shield (copper), added along the slab direction, 
leads to an estimation of the temperature gradient along the slab of around ${\rm 7^o C}$. This simulation is probably 
pessimistic since only copper is used and not the other material of the slab (PCB, tungsten, carbon fibers...), 
but the latest simulations are correlated with tests. Therefore, passive cooling inside the slab could be 
sufficient for the full scale ECAL. Then, the main cooling system could be at the end of each slab. 

\begin{figure}
\centerline{\includegraphics[width=0.9\columnwidth]{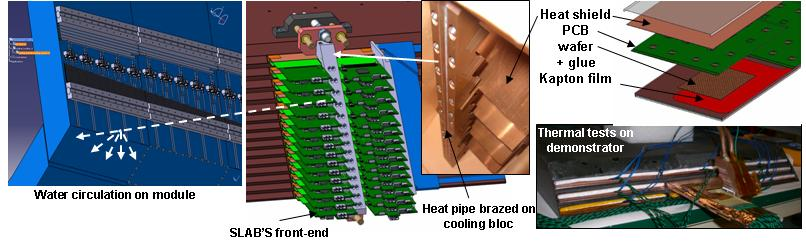}}
\caption{Possible cooling system for ECAL barrel and end-caps}
\label{fig:eudet_thermal}
\end{figure}

For each module, on the front end, cooling pipes connected to each heat shield, should be located in the gap 
between ECAL and HCAL, with all slabs' input/output for the readout and power supply. This front slab thermal 
connection has been designed and tested in order to allow safe contact and the best heat conduction with heat shields. 
Total power to dissipate will be approximately 4565 W for about 82,2 M of channels for the whole ECAL. 
After thermal study, design and tests of heat pipes technology are on going, testing connection to slab and 
prototype of full system for 1 slab and 1 column.  The thermal simulations show that the thermal flux through 
the module is an important data to be correlated by tests (on EUDET module after the last Autumn 2009 thermal 
test with the Demonstrator) to know thermal influence of ECAL on neighboring detectors.

The study and design of the global cooling system for ECAL including the power source, fluid lines, valves, sensors
and safety systems allows a first cost estimate and the various options of the whole installation. Indeed, the global 
system with temperature and power range adapted can be designed with a leak less system including or not heat pipe 
termination in order to prevent front-end electronic from any water sparkling. 
In this way, a true scale leak less cooling system test has been performed in 2009 with representative systems to 
control, correlating the simulation of such a system; a real loop will be constructed and tested in 2010 to validate 
process, regulation, interface and control.
A 3D pipe modeling of the whole ECAL equipped should give important information for detector integration.

\subsection{Data Acquisition system}

The DAQ system envisaged for the EUDET ECAL prototype is based on the CALICE second generation DAQ. This has
been designed with a view to use in a full scale detector: it is highly modular, and is based on off the shelf
hardware. The Detector Interface card (DIF), currently the size of a credit card, will be placed at the end of each
detector slab, in the gap between the ECAL and HCAL. It is the only detector-specific element of the system, the
other elements are common between all systems. The signals from several DIFs are collected by the 
Local Data Aggregation card (LDA), which passed the data out of the detector via an optical link.
A Clock and Control Card (CCC) distributes clock signals to the LDAs.
The hardware for this system is available, and the firmware and software required is currently under development.

\section{Interface with the HCAL: mounting and alignment}

\begin{figure}
\centerline{\includegraphics[width=0.9\columnwidth]{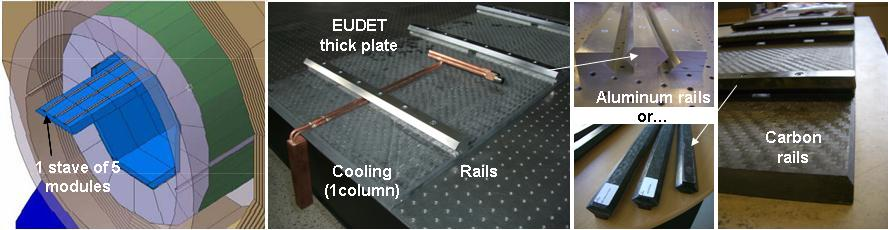}}
\caption{Barrel and endcap ECAL/HCAL interface with its fastening and guiding system}
\label{fig:fixing}
\end{figure}

The alignment and cooling system are both crucial to ensure the needed detector performances from the point of 
view of correct operation and good precision. The ECAL will be fastened to the HCAL stave 
by stave, through accurate rail systems. The main issue designing this rail system is to reduce as small as 
possible the gap between ECAL and HCAL to keep the continuity between the 2 calorimeters (typically less than 4 cm). 
However, this gap has to be sufficient to install all the different services like cooling pipes and fans, 
electrical power and signal distribution. A fastening system, based on rails which will be fixed by metal inserts, 
directly inside the composite structure of the ECAL could solve this problem 
(see Figure \ref{fig:fixing}). But this gap has to be optimized according to mechanical simulations and destructive tests of the 
ECAL/HCAL interface and the needs of cooling and fluid systems.

%

\section{Conclusions}

The CALICE second generation SiW ECAL prototype is under development. 
It will provide essential technical input to the construction of modules of a future SiW ECAL at an ILC detector. 
Mechanical prototypes have allowed the testing of manufacturing techniques, 
and have been tested to provide input to mechanical simulations of the detector modules. 
Thermal tests of the same prototypes have allowed their thermal characteristics to be well understood.

The silicon sensors are the cost drivers of this detector, and an effort is underway to understand
strategies to reduce the cost in the context of a large-scale production. The integration of the
sensors and FE electronics into the detector are the subject of studies into thin PCB
manufacturing and wire bonding of unpackaged chips into the PCBs. Tools for the
integration of the various elements have been developed.

A scalable DAQ system is being developed, which will be used to run future
cosmic and beam tests of the prototype and its constituents as they are built.

The mechanical and thermal issues of the integration of the ECAL into a future detector
are also under study.

\bibliographystyle{plain}
\bibliography{myrefs}

\end{document}